\documentstyle[epsfig,12pt]{article}
\begin {document}

\title {CONFORMALLY FLAT ANISOTROPIC SPHERES IN GENERAL RELATIVITY.}

\author{L. Herrera\thanks{ Also at UCV, Caracas, Venezuela; e-mail address:
lherrera@gugu.usal.es}
 , A. Di Prisco\thanks{On leave from Universidad Central de Venezuela,
Caracas, Venezuela}, J. Ospino\\
Area de F\'\i
sica Te\'orica. Facultad de Ciencias.\\ Universidad de Salamanca. 37008
Salamanca, Espa\~na.\\
and\\
E. Fuenmayor\\
Escuela de F\'\i sica, Facultad de Ciencias,\\
Universidad Central de Venezuela,\\
Caracas, Venezuela.
}

\date{}
\maketitle

\begin{abstract}
The condition for the vanishing of the Weyl tensor is integrated in
the spherically symmetric case. Then, the resulting expression is
used to find new, conformally flat, interior solutions to Einstein
equations for locally anisotropic fluids. The slow evolution of these
models is contrasted with the evolution of models with similar energy
density or radial pressure distribution but non-vanishing Weyl tensor,
thereby bringing out the different role played by the Weyl tensor,
the local anisotropy of pressure and the inhomogeneity of the
energy density in the collapse of relativistic spheres.
\end{abstract}
\newpage

\section{Introduction}
In the study of self-gravitating systems there are three factors
whose relevance has been recurrently and separately stressed in
the literature.
These are, the Weyl tensor, the local anisotropy of pressure and
the inhomogeneity of energy density distribution (density contrast).

The Weyl tensor \cite{Pe} or some functions of it \cite{Wa}, have
been proposed to provide a gravitational arrow of time. The rationale
behind this idea being that tidal forces tend to make the gravitating
fluid more inhomogeneous as the evolution proceeds, thereby
indicating the sense of time. However, some works have thrown
doubts on this proposal \cite{Bo}. Also, as it will be seen below,
it is worth noticing that the relation between the Weyl tensor
and the density contrast is affected by the presence of local
anisotropy of pressure.

The role of density inhomogeneities in the collapse of dust \cite{MeTa}
and in particular in the formation of naked singularities \cite{VarI},
has been extensively discussed in the literature.

Finally, the assumption of local anisotropy of pressure, has been proved
to be very useful in the study of relativistic compact objects
(see \cite{HeSa97} and references therein).

A hint pointing to the relevance of the above mentioned three factors in
the fate of spherical collapse is also provided by the expression of
the active gravitational mass in terms of those factors
\cite{Hetal}, \cite{HeSa95}.

These three factors are usually considered separately,
their relationship being omitted from discussion,even though they are
related by a simple expression, which we shall present below
\cite{Hetal}, \cite{HeSa95}.

The purpose of this work is twofold. On one hand we shall integrate
the vanishing Weyl tensor condition, which will allow us to construct
conformally flat models (with anisotropic pressure). The obtained
solutions represent static or slowly evolving (in the quasi-static
approximation) spheres, which could serve for the modelling of compact
self-gravitating objects. On the other hand, we want to study,
comparatively, the effects of the above mentioned parameters on the (slow)
evolution of relativistic spheres. For doing this we shall contrast the
evolution of the conformally flat models with the evolution of
models with the same energy density or radial pressure distribution,
but non-vanishing Weyl tensor. With this purpose it will be useful to
calculate the active gravitational
mass and the fluid velocity for each model.

The paper is organized as follows: In the next Section all relevant
equations and conventions are given. The condition for the vanishing
of the Weyl tensor is integrated in Section 3 and the models are
described in Section 4. Finally a discussion of results is presented
in the last Section.

\section{Relevant equations and conventions}
\subsection{The field equations}
We consider a spherically symmetric distribution of collapsing
fluid, which we assume to be locally anisotropic and
bounded by a
spherical surface $\Sigma$.
The line element is given in Schwarzschild-like coordinates by
 \begin{equation}
ds^2=e^{\nu} dt^2 - e^{\lambda} dr^2 -
r^2 \left( d\theta^2 + sin^2\theta d\phi^2 \right)
\label{metric}
\end{equation}
where $\nu$ and $\lambda$ are functions of $t$ and $r$.
The coordinates are: $x^0=t; \, x^1=r; \, x^2=\theta; \, x^3=\phi$.

The metric (\ref{metric}) has to satisfy Einstein field equations
\begin{equation}
G^\mu_\nu=-8\pi T^\mu_\nu
\label{Efeq}
\end{equation}
which in our case read \cite{Bd}:
\begin{equation}
-8\pi T^0_0=-\frac{1}{r^2}+e^{-\lambda}
\left(\frac{1}{r^2}-\frac{\lambda'}{r} \right)
\label{feq00}
\end{equation}
\begin{equation}
-8\pi T^1_1=-\frac{1}{r^2}+e^{-\lambda}
\left(\frac{1}{r^2}+\frac{\nu'}{r}\right)
\label{feq11}
\end{equation}
\begin{eqnarray}
-8\pi T^2_2  =  -  8\pi T^3_3 = & - &\frac{e^{-\nu}}{4}\left(2\ddot\lambda+
\dot\lambda(\dot\lambda-\dot\nu)\right) \nonumber \\
& + & \frac{e^{-\lambda}}{4}
\left(2\nu''+\nu'^2 -
\lambda'\nu' + 2\frac{\nu' - \lambda'}{r}\right)
\label{feq2233}
\end{eqnarray}
\begin{equation}
-8\pi T_{01}=-\frac{\dot\lambda}{r}
\label{feq01}
\end{equation}
where dots and primes stand for partial differentiation with respect
to $t$ and $r$
respectively.

In order to give physical significance to the $T^{\mu}_{\nu}$ components
we apply the Bondi approach \cite{Bd},
i.e we introduce  local Minkowski
coordinates ($\tau, x, y, z$), defined by
$$d\tau=e^{\nu/2}dt\,\qquad\,dx=e^{\lambda/2}dr\,\qquad\,
dy=rd\theta\,\qquad\, dz=rsin\theta d\phi$$
Then, denoting the Minkowski components of the energy tensor by a bar,
we have
$$\bar T^0_0=T^0_0\,\qquad\,
\bar T^1_1=T^1_1\,\qquad\,\bar T^2_2=T^2_2\,\qquad\,
\bar T^3_3=T^3_3\,\qquad\,\bar T_{01}=e^{-(\nu+\lambda)/2}T_{01}$$
Next we suppose that, when viewed by an observer moving relative to these
coordinates with velocity $\omega$ in the radial direction, the physical
content  of space consists of an anisotropic fluid of energy density $\rho$,
radial pressure $P_r$ and tangential pressure $P_\bot$.
Thus, when viewed by this moving observer, the covariant energy-momentum
tensor in
Minkowski coordinates is

\[ \left(\begin{array}{cccc}
\rho    &  0        &   0     &   0    \\
0       &  P_r      &   0     &   0    \\
0       &   0       & P_\bot  &   0    \\
0       &   0       &   0     &   P_\bot
\end{array} \right) \]
Then a Lorentz transformation readily shows that
\begin{equation}
T^0_0=\bar T^0_0= \frac{\rho + P_r \omega^2 }{1 - \omega^2}
\label{T00}
\end{equation}
\begin{equation}
T^1_1=\bar T^1_1=-\frac{ P_r + \rho \omega^2}{1 - \omega^2}
\label{T11}
\end{equation}
\begin{equation}
T^2_2=T^3_3=\bar T^2_2=\bar T^3_3=-P_\bot
\label{T2233}
\end{equation}
\begin{equation}
T_{01}=e^{(\nu + \lambda)/2} \bar T_{01}=
-\frac{(\rho + P_r) \omega e^{(\nu + \lambda)/2}}{1 - \omega^2}
\label{T01}
\end{equation}
Note that the velocity in the ($t,r,\theta,\phi$) system, $dr/dt$,
is related to $\omega$ by
\begin{equation}
\omega=\frac{dr}{dt}\,e^{(\lambda-\nu)/2}
\label{omega}
\end{equation}

Outside of the fluid, the spacetime is
Schwarzschild,
\begin{equation}
ds^2= \left(1-\frac{2M}{r}\right) dt^2
- \left(1-\frac{2M}{r}\right)^{-1} dr^2
- r^2 \left(d\theta^2 + sin^2\theta d\phi^2 \right)
\label{Sch}
\end{equation}
In order to match the two metrics smoothly   on the boundary surface
$r=r_\Sigma(t)$, we  require  continuity of the first
and second fundamental
forms across that surface. As result of this matching we obtain
the well known result
\begin{equation}
\left[P_r\right]_\Sigma = 0
\label{Psig}
\end{equation}

Next,the radial component of the
conservation law
\begin{equation}
T^\mu_{\nu;\mu}=0
\label{dTmn}
\end{equation}
gives
\begin{equation}
\left(-8\pi T^1_1\right)'=\frac{16\pi}{r} \left(T^1_1-T^2_2\right)
+ 4\pi \nu' \left(T^1_1-T^0_0\right) +
\frac{e^{-\nu}}{r} \left(\ddot\lambda + \frac{\dot\lambda^2}{2}
- \frac{\dot\lambda \dot\nu}{2}\right)
\label{T1p}
\end{equation}
which in the static case becomes
\begin{equation}
P'_r=-\frac{\nu'}{2}\left(\rho+P_r\right)+
\frac{2\left(P_\bot-P_r\right)}{r}
\label{Prp}
\end{equation}
representing the generalization of the Tolman-Oppenheimer-Volkof equation
for anisotropic fluids \cite{HeSa97}.

In this work we shall consider exclusively static or slowly evolving
(quasi-static) systems. By this we mean that our sphere either does not change
or changes slowly on a time scale that is very long compared to the typical
time in which the sphere reacts to a slight perturbation of hydrostatic
equilibrium (this typical time scale is called hydrostatic time scale).
Thus our system is always very close to or in hydrostatic
equilibrium and its evolution may be regarded as a sequence of static
models linked by (\ref{feq01}). This assumption is very sensible because
the hydrostatic
time scale is very small for almost any phase of the life of the star.
It is of the order of $27$ minutes for the Sun, $4.5$ seconds for a white dwarf
and $10^{-4}$ seconds for a neutron star of one solar mass and $10$ Km radius.
It is well known that any of the stellar configurations mentioned above,
change on a time scale that is very long compared to their respective
hydrostatic time scales. Let us now translate this assumption in conditions
to $\omega$ and metric functions.

First of all, slow contraction means that the radial velocity $\omega$ as
measured by the Minkowski observer is always much smaller than the velocity
of light ($\omega \ll 1$). Therefore we shall neglect terms of the order
$O(\omega^2)$.

Then (\ref{T1p}) yields
\begin{equation}
\ddot\lambda + \frac{\dot\lambda^2}{2} -
\frac{\dot\nu \dot\lambda}{2} = 8 \pi r e^{\nu}
\left[P'_r + \left(\rho + P_r\right) \frac{\nu'}{2} -
2 \frac{P_\bot - P_r}{r}\right]
\label{lsps}
\end{equation}

Since by assumption, our system is always (not only at a given time $t$)
in equilibrium (or very close to), (\ref{Prp}) and (\ref{lsps}) imply then,
for an arbitrary slowly evolving configuration
\begin{equation}
\ddot\lambda \approx \dot\nu \dot\lambda \approx
\dot\lambda^2 \approx 0
\label{lp0}
\end{equation}
and of course, time derivatives of any order of the left hand side of the
hydrostatic equilibrium equation must also vanish, for otherwise the
system will deviate from equilibrium. This condition implies, in particular,
that we must demand
$$
\ddot\nu \approx 0
$$

Finally, from the time derivative of (\ref{feq01}), and using (\ref{T01}),
it follows that

\begin{equation}
\dot\omega \approx O(\ddot\lambda, \dot\lambda \omega, \dot\nu \omega)
\label{Omo}
\end{equation}
which implies that we shall also neglect terms linear in the acceleration.
On purely physical considerations, it is obvious that the vanishing of
$\dot\omega$ is required to keep the system always in equilibrium.

Thus, from now on, we shall always assume
\begin{equation}
O(\omega^2) = {\dot\lambda}^2 = {\dot\nu}^2 =
\dot\lambda \dot\nu = \ddot\lambda = \ddot\nu = 0
\label{om2}
\end{equation}
implying that the system remains in (or very close to) equilibrium.

\subsection{The Weyl tensor}
We can now calculate the components of the Weyl tensor.
Neglecting terms of order $\dot\lambda \dot\nu, {\dot\lambda}^2,
{\dot\nu}^2, \ddot\lambda$ and $\ddot\nu$, we find that all
non-vanishing components can be expressed through $C^3_{232}$.
Thus,
\begin{equation}
W \equiv \frac{r}{2} C^3_{232} =
\frac{r^3 e^{-\lambda}}{6} \left(
\frac{e^\lambda}{r^2} + \frac{\nu' \lambda'}{4} -
\frac{1}{r^2} - \frac{\nu'^2}{4} - \frac{\nu''}{2} -
\frac{\lambda' - \nu'}{2r}
\right)
\label{W}
\end{equation}

Next, defining the mass function as usual
\begin{equation}
m(r,t) = 4 \pi \int^r_0{T^0_0 r^2 dr}
\label{mf}
\end{equation}
the following relations may be established (\cite{Hetal},\cite{HeSa95})
\begin{equation}
W = - \frac{4}{3} \pi \int_0^r{r^3 \left(T^0_0\right)' dr} +
\frac{4}{3} \pi r^3 \left(T^2_2 - T^1_1\right)
\label{Wint}
\end{equation}
\begin{equation}
m(r,t) = \frac{4}{3} \pi r^3 T^0_0 -
\frac{4}{3} \pi \int_0^r{r^3 \left(T^0_0\right)' dr}
\label{mint}
\end{equation}
Both, (\ref{Wint}) and (\ref{mint}) are valid in the general (dynamic) case.
However only in the static or the quasi-static case
$T^0_0$ and $T^1_1$ denote the
proper energy density and the radial pressure respectively.

\subsection{The Tolman-Whittaker mass}
The Tolman-Whittaker mass \cite{ToWi} within a sphere of radius $r$
inside $\Sigma$, is defined as \cite{HeSa95}
\begin{equation}
m_{TW}(r,t) = 4 \pi \int_0^{r}
{r^2 e^{\left(\nu + \lambda\right)/2}
\left(T^0_0 - T^1_1 - 2 T^2_2\right) dr}
\label{mTW}
\end{equation}

Two alternative expressions, easily obtained from the field
equations (see \cite{HeSa95} for details) are
\begin{equation}
m_{TW} = e^{\left(\nu + \lambda\right)/2}
\left(m + 4 \pi P_r r^3\right)
\label{mtwm}
\end{equation}
and
\begin{equation}
m_{TW} = \frac{1}{2} e^{\left(\nu - \lambda\right)/2} \nu' r^2
\label{mnup}
\end{equation}

The interpretation of $m_{TW}$ as the active gravitational mass
follows at once from (\ref{mnup}) and (\ref{Prp}). Indeed, the
first term on the right side of (\ref{Prp}) (the ``gravitational force''
term) is the product of the passive gravitational mass density
($\rho + P_r$) and a term proportional to $m_{TW}/r^2$. A similar
conclusion can be obtained if we recall that the gravitational
acceleration of a test particle, instantaneously at rest in a
static gravitational field, as measured with standard rods and
coordinate clocks is given by \cite{Gr}
\begin{equation}
a= - \frac{e^{(\nu-\lambda)/2}\nu'}{2} = - \frac{m_{TW}}{r^2}
\label{a}
\end{equation}

\subsection{The velocity of a fluid element}
For the comparative study of the (slow) evolution of different solutions,
it will be useful to plot the velocity ($\omega$) profiles for
different pieces of material. A simple expression for $\omega$, may
be obtained as follows: from (\ref{feq00}) and (\ref{mf}), it results
\begin{equation}
e^{-\lambda} = 1-\frac{2m}{r}
\label{el}
\end{equation}
and from (\ref{el})
\begin{equation}
\dot m = \frac{\dot\lambda r e^{-\lambda}}{2}
\label{mp}
\end{equation}
then using (\ref{feq01})
\begin{equation}
\omega = - \frac{\dot m e^{(\lambda-\nu)/2}}{4 \pi r^2 (\rho + P_r)}
\label{om}
\end{equation}

\section{The vanishing Weyl condition}
We shall now proceed to integrate the condition
\begin{equation}
W=0
\label{W0}
\end{equation}
which, using (\ref{W}), may be written as
\begin{equation}
\left(\frac{e^{-\lambda}\nu'}{2r}\right)'
+ e^{-(\nu+\lambda)} \left(\frac{e^{\nu}\nu'}{2r}\right)' -
\left(\frac{e^{-\lambda}-1}{r^2}\right)' = 0
\label{pr}
\end{equation}
Introducing new variables
\begin{equation}
y = e^{-\lambda} \qquad ; \qquad \frac{\nu'}{2} = \frac{u'}{u}
\label{cv}
\end{equation}
equation(\ref{pr}) is cast into
\begin{equation}
y' +
\frac{2y \left[u''- u'/r + u/r^2 \right]}{\left[u' - u/r\right]} -
\frac{2u}{r^2 \left[u' - u/r\right]} = 0
\label{yp}
\end{equation}
whose formal solution is
\begin{equation}
y = e^{-\int{k(r)dr}} \left[\int{e^{\int{k(r)dr}}f(r)dr}+ C_1\right]
\label{y}
\end{equation}
where $C_1$ is a constant of integration, and
\begin{equation}
k(r) = 2 \frac{d}{dr}\left[\ln{\left(u'- \frac{u}{r}\right)}\right]
\label{g}
\end{equation}
\begin{equation}
f(r) = \frac{2u}{r^2 \left(u'- u/r\right)}
\label{f}
\end{equation}
changing back to the original variables, eq.(\ref{y}) becomes
\begin{equation}
\frac{\nu'}{2} - \frac{1}{r} = \frac{e^{\lambda/2}}{r}
\sqrt{1-c^2 r^2 e^{-\nu}}
\label{1}
\end{equation}
with $c^2 \equiv - C_1$.

Next, (\ref{1}) may be formally integrated, to obtain
\begin{equation}
e^\nu = c^2 r^2 \cosh^2{\left[\int{\frac{e^{\lambda/2}}{r}dr}+ \tilde{C}\right]}
\label{fint}
\end{equation}
where $\tilde{C}$ is a constant of integration (a function of $t$, in the
slowly evolving case). The reader may check that
(\ref{fint}) satisfies (\ref{1}) and (\ref{W0}) (or (\ref{pr})).

In the next section we shall present some models satisfying (\ref{1})
(or (\ref{fint})).

\section{The models}
A simple counting of equations ((\ref{feq00})--(\ref{feq01})) and
unknowns ($\nu, \lambda, \rho, P_r, P_\bot, \omega$) indicates that we have to
provide two additional relations (in the form of equations of state and/or
restrictions on metric variables), in order to integrate the system
((\ref{feq00})--(\ref{feq01})).

If one assumes that the fluid is locally isotropic ($P_r=P_\bot$) then,
demanding $W=0$, we are driven to a unique solution (the Schwarzschild
interior solution), a fact also obvious from (\ref{Wint}). However, if
$P_r\not=P_\bot$,
then the condition $W=0$, does not single out a unique model.

In what follows we shall construct two models with $W=0$. One of them
characterized by $P_r=0$, and for the other we prescribe a given energy
density distribution which is similar to the one proposed by Gokhroo and
Mehra \cite{GM}. Additionally, we present two other models with
$W\not=0$. One is characterized by $P_r=0$ and $\rho=\rho(t)$
(\cite{HeSa95},\cite{Fl}), and the other has the same energy density
distribution as one of the conformally flat solutions.

\subsection{Model I}
Our first model is characterized by
\begin{equation}
W=0
\label{WI}
\end{equation}
and
\begin{equation}
P_r=0
\label{prI}
\end{equation}

Then, from (\ref{feq11}) and (\ref{1}), it follows
\begin{equation}
e^{-\nu}= \frac{g}{c^2 r^2} \, \frac{(4-9g)}{(1-2g)}
\label{nuI}
\end{equation}
with
\begin{equation}
g\equiv \frac{m(r,t)}{r}
\label{gI}
\end{equation}
and where (\ref{el}) has been used.

Next, taking $r$-derivative of (\ref{nuI}), and using
\begin{equation}
\nu'= \frac{2m}{r^2 \left(1-2m/r\right)} \equiv
\frac{2g}{r(1-2g)}
\label{nupI}
\end{equation}
easily derived from (\ref{feq11}), we obtain
\begin{equation}
g'= \frac{54g^3 - 42g^2 + 8g}{r \left(18g^2 - 18g + 4\right)}
\label{gpI}
\end{equation}
which after integration yields
\begin{equation}
D r = \frac{g^{1/2}}{(4-9g)^{1/6}}
\label{ctr}
\end{equation}
where $D$ is a constant (a function of $t$ in the slowly
evolving case) of integration. Solving (\ref{ctr}) for $g$, one obtains
\begin{equation}
g = a^{1/3} \left\{
\left[2 + \left(27a + 4\right)^{1/2}\right]^{1/3}
+ \left[2 - \left(27a + 4\right)^{1/2}\right]^{1/3}
\right\}
\label{gfI}
\end{equation}
with
\begin{equation}
(D r)^6 = a \equiv \left(\frac{r}{r_\Sigma}\right)^6
\frac{g_\Sigma^3}{4-9g_\Sigma}
\label{aI}
\end{equation}
where subscript $\Sigma$ indicates that the quantity is evaluated
at the boundary surface $r=r_\Sigma$.

The remaining variables are now easily obtained from the field equations
and (\ref{nuI}), (\ref{gfI}). Thus,
\begin{equation}
\rho = \frac{3g}{2 \pi r^2} \, \frac{(1-2g)}{(2-3g)}
\label{roI}
\end{equation}
\begin{equation}
P_\bot = \frac{3g^2}{4 \pi r^2} \, \frac{1}{(2-3g)}
\label{ptI}
\end{equation}
\begin{equation}
e^{-\lambda} = 1 - 2g
\label{lI}
\end{equation}

For the Tolman-Whittaker mass we obtain, using either (\ref{mtwm})
or (\ref{mnup})
\begin{equation}
m_{TW} = g \left(\frac{r}{r_\Sigma}\right) r \left[
\frac{g_\Sigma \left(1-\frac{9}{4}g_\Sigma\right)}{g
\left(1-\frac{9}{4}g\right)}
\right]^{1/2}
\label{mtwI}
\end{equation}
or, using the dimensionless varibles
\begin{equation}
x \equiv \frac{r}{r_\Sigma} \qquad ; \qquad
n = \frac{m_\Sigma}{r_\Sigma} \equiv \frac{M}{r_\Sigma} = g_\Sigma
\label{sdI}
\end{equation}
\begin{equation}
m_{TW} = \frac{M x^3 \left(Z_1 + Z_2\right)^{1/2} \left(4-9n\right)^{1/2}}
{\left[4 \left(4-9n\right)^{1/2} - 9 n x^2 \left(Z_1 + Z_2\right)\right]^{1/2}}
\label{admI}
\end{equation}
with
\begin{equation}
Z_{1,2} = \left[2 \left(4-9n\right)^{1/2} \pm \left(
27 x^6 n^3 + 16 - 36n\right)^{1/2}\right]^{1/3}
\label{ZI}
\end{equation}

It is worth noticing that from the requirement $\rho \geq 0$, it follows,
using (\ref{roI})
\begin{equation}
g<\frac{2}{3}
\label{cog}
\end{equation}
a stronger restriction appears from the condition
\begin{equation}
\rho \geq P_\bot
\label{sc}
\end{equation}
which requires
\begin{equation}
g<\frac{2}{5}
\label{cogs}
\end{equation}

Finally, the velocity $\omega$ of any fluid element is given by
\begin{eqnarray}
\omega = && \frac{\omega_\Sigma n^{1/2} x^2 (2-3g) (4-9g)^{1/2} (1-2n)}
{g^{1/2} (4-9n)^2 (1-2g)^2} \times \nonumber \\
&\times& \left\{2 \left(Z_1 + Z_2\right) +
\frac{3 n x^2 \left(Z_2^{2} - Z_1^{2}\right)}{\left(27x^6n^3 + 16 -
36n\right)^{1/2}}\right\}
\label{omI}
\end{eqnarray}
where (\ref{omega}), (\ref{om}), (\ref{gfI})--(\ref{roI}),
(\ref{sdI}), and (\ref{ZI}) have been used.

\subsection{Model II}
The second model we shall consider, is well known. It is characterized by
\begin{equation}
P_r = 0
\label{prII}
\end{equation}
\begin{equation}
\rho = \rho(t)
\label{rot}
\end{equation}
and is not conformally flat.

A static version of this model was studied by Florides \cite{Fl}.
Models with vanishing radial pressure have been discussed in the past
\cite{VarII}, and more recently in relation with the formation
of naked singularities \cite{Ma}.

The corresponding variables are (see \cite{HeSa95} for details)
\begin{equation}
\rho = \rho(t) \qquad ; \qquad P_r = 0
\label{roprII}
\end{equation}
\begin{equation}
P_\bot = \frac{2 \pi r^2 \rho^2}{3 \left(1 - \frac{8 \pi}{3} r^2 \rho\right)}
\label{ptII}
\end{equation}
\begin{equation}
e^\nu = \frac{\left(1 - \frac{8 \pi}{3} r_\Sigma^2 \rho\right)^{3/2}}
{\left(1 - \frac{8 \pi}{3} r^2 \rho\right)^{1/2}}
\label{nuII}
\end{equation}
\begin{equation}
e^{-\lambda} = 1 - \frac{8 \pi}{3} r^2 \rho
\label{lII}
\end{equation}
\begin{equation}
W = - \frac{8 \pi^2 r^5 \rho^2}{9 \left(1 - \frac{8 \pi}{3} r^2 \rho\right)}
\label{WII}
\end{equation}
\begin{equation}
m_{TW}= \frac{4 \pi}{3} r^3 \rho
\left(\frac{1 - \frac{8 \pi}{3} r_\Sigma^2 \rho}
{1 - \frac{8 \pi}{3} r^2 \rho}\right)^{3/4}
\label{mII}
\end{equation}
\begin{equation}
\omega= \omega_\Sigma \left(\frac{r}{r_\Sigma}\right)
\left(\frac{1 - \frac{8 \pi}{3} r_\Sigma^2 \rho}
{1 - \frac{8 \pi}{3} r^2 \rho}\right)^{1/4}
\label{omII}
\end{equation}

\subsection{Model III}
This model is conformally flat and is further characterized by
\begin{equation}
e^{-\lambda} = \left(1 - \frac{r^2}{b^2}\right)^{2}
\label{lIII}
\end{equation}
where $b$ is a constant (a function of $t$ in the slowly evolving case).

Then, from (\ref{fint}) one obtains
\begin{equation}
e^{\nu} = \frac{c^2}{4 B^2}
\frac{\left[B^2 r^2 + b^2 - r^2\right]^2}{\left(b^2 - r^2\right)}
\label{nuIII}
\end{equation}
with
\begin{equation}
\tilde{C} = \ln{B}
\label{Ctt}
\end{equation}

>From the field equations, (\ref{lIII}) and (\ref{nuIII}), we can now
obtain the expressions for $\rho$, $P_r$ and $P_\bot$.

\begin{equation}
\rho = \frac{3}{4 \pi b^2} \left(1 - \frac{5 r^2}{6 b^2}\right)
\label{roIII}
\end{equation}
\begin{equation}
8 \pi P_r = - \frac{1}{r^2} + \frac{(b^2-r^2)^{2}}{b^4}
\left[
\frac{1}{r^2} +
\frac{4 {B}^2 b^2 - 2 [{B}^2 r^2 +
b^2-r^2]}
{(b^2-r^2)[{B}^2 r^2 + b^2-r^2]}
\right]
\label{prIII}
\end{equation}
\begin{equation}
8 \pi \left(P_r - P_\bot\right)= - \frac{2 r^2}{b^4}
\label{anis}
\end{equation}
$b$, $c$ and $B$ are related to the total
mass and the radius of the sphere through the boundary conditions
\begin{equation}
e^{-\lambda_\Sigma} = 1 - \frac{2M}{r_\Sigma}
\label{lsIII}
\end{equation}
\begin{equation}
 e^{\nu_\Sigma} = 1 - \frac{2M}{r_\Sigma}
\label{nusIII}
\end{equation}
\begin{equation}
P_{r_\Sigma} = 0
\label{pr0III}
\end{equation}
the corresponding expressions are
\begin{equation}
b = \frac{r_\Sigma}{\left[1 - \left(1 -
\frac{2M}{r_\Sigma}\right)^{1/2}\right]^{1/2}}
\label{bIII}
\end{equation}
\begin{equation}
c = \frac{1}{r_\Sigma}
\left[\frac{4M}{r_\Sigma} - \frac{9M^2}{r_\Sigma^2}\right]^{1/2}
\label{cIII}
\end{equation}
\begin{equation}
B =
\frac{\left(1-2M/r_\Sigma\right)^{1/4}
\left[\left(1-2M/r_\Sigma\right)^{1/2}+
\left(3M/r_\Sigma - 1\right)\right]}{\left[1
- \left(1 - 2M/r_\Sigma\right)^{1/2}\right]^{1/2}
\left[4M/r_\Sigma - 9M^2/r_\Sigma^2\right]^{1/2}}
\label{CTT}
\end{equation}

For the Tolman-Whittaker mass the obtained expression is
\begin{equation}
m_{TW} = \frac{c r^3}{2 {B} b^2 (b^2-r^2)^{1/2}}
\left[2 {B}^2 b^2 - {B}^2 r^2 -
b^2+r^2\right]
\label{mTWIII}
\end{equation}
or, using (\ref{bIII})--(\ref{CTT}) and (\ref{sdI})
\begin{eqnarray}
& &m_{TW}= \frac{ M x^3 (1-2n)^{1/4}}{\left[1-x^2 + x^2
(1-2n)^{1/2}\right]^{1/2}} \times \\
&&\left\{3 - 2x^2 - \frac{(1-x^2)}{n}\left[1-(1-2n)^{1/2}\right]
\left[1 + \frac{4n-9n^2}{2(1-2n)^{1/2}\left[(1-2n)^{1/2}
+ 3n-1\right]}\right]\right\} \nonumber
\label{mTWf}
\end{eqnarray}
and the expression for $\omega$ in this model, results in
\begin{eqnarray}
\omega =&& \frac{4 \dot{b}{B} x
\left[1 - (1-2n)^{1/2}\right]}{\left(4n-9n^2\right)^{1/2}
\left(1-x^2 \left[1-(1-2n)^{1/2}\right]\right)^{1/2}} \times \\
&\times&
\frac{1}{\left[2 {B}^{2} +
{B}^{2} x^2 \left[1 - (1-2n)^{1/2}\right] +
1 - x^2 \left[1 - (1-2n)^{1/2}\right]\right]} \nonumber
\label{omIII}
\end{eqnarray}
where $\dot{b}$ is easily obtained from (\ref{bIII})
\begin{equation}
\dot{b} = \frac{\omega_\Sigma n (1-2n)^{1/2}
\left(5 - \frac{2}{n} \left[1-(1-2n)^{1/2}\right]\right)}
{2 \left[1 - (1-2n)^{1/2}\right]^{3/2}}
\label{bpun}
\end{equation}

\subsection{Model IV}
This model has the same energy density distribution as the previous one
(same $\lambda$), but is not conformally flat ($W\not=0$).
Instead, the model is further characterized by
\begin{equation}
P_r=0
\label{prce}
\end{equation}

Then, from (\ref{lIII}), (\ref{prce}) and
field equations, we obtain
\begin{equation}
\rho = \frac{3}{4 \pi b^2} \left(1 - \frac{5 r^2}{6 b^2}\right)
\label{roIV}
\end{equation}
\begin{equation}
e^\nu = \frac{\beta}{\left(b^2 -r^2\right)^{1/2}}
e^{\frac{b^2}{2 \left(b^2-r^2\right)}}
\label{nuIV}
\end{equation}
\begin{equation}
P_\bot = \frac{\left(2b^2r^2-r^4\right)\left(6b^2-5r^2\right)}
{32 \pi b^4 \left(b^2-r^2\right)^2}
\label{PtIV}
\end{equation}
with
\begin{equation}
\beta \equiv b \left(1 - \frac{2M}{r_\Sigma}\right)^{5/4}
e^{-\frac{1}{2\left(1-2M/r_\Sigma\right)^{1/2}}}
\label{be}
\end{equation}
and $b$ is given by (\ref{bIII}).

For the active gravitational mass, we obtain, after some lengthly
calculations
\begin{eqnarray}
m_{TW} &=& M (1-2n)^{5/8} x^3 \left(1 - (1-2n)^{1/2}\right)
\frac{\left[1 - \frac{x^2}{2} \left(1 - (1-2n)^{1/2}\right)\right]}
{n \left(1-x^2 \left[1-(1-2n)^{1/2}\right]\right)^{5/4}} \times
\nonumber \\
&\times&\exp{\frac{1}{4}\left[\frac{(x^2-1)\left(1 - (1-2n)^{1/2}\right)}
{(1-2n)^{1/2}\left[1-x^2 \left(1 - (1-2n)^{1/2}\right)\right]}\right]}
\label{mIV}
\end{eqnarray}
where $n$ and $x$ are defined by eq.(\ref{sdI}). Observe that condition
$\rho \geq 0$ is satisfied for all $n$, however if we demand $\rho \geq
P_\bot$,
then
\begin{equation}
n \leq 0.4
\label{n}
\end{equation}

Finally, the expression for the velocity takes the form

\begin{eqnarray}
\omega &=& \frac{2 \omega_\Sigma x \left[1 - x^2 + x^2 (1-2n)^{1/2}\right]^{1/4}
\left[5n - 2 + 2 (1-2n)^{1/2}\right] }
{(1-2n)^{1/8} \left[6 - 5 x^2 + 5 x^2 (1-2n)^{1/2}\right]
\left[1 - (1-2n)^{1/2}\right]} \times \nonumber \\
&&\exp{\frac{1}{4} \left[\frac{(1 - x^2) \left[1 - (1-2n)^{1/2}\right]}
{(1-2n)^{1/2} \left[1 - x^2 + x^2 (1-2n)^{1/2}\right]}\right]}
\label{omIV}
\end{eqnarray}

\section{Discussion}
We have integrated the vanishing Weyl condition. The resulting
expression (\ref{1}) (and (\ref{fint})) allows to find conformally
flat models in a very simple way, once an additional condition on
physical or metric variables is imposed.
Specifically we have found two conformally flat models (I, III).
In order to bring out the role of Weyl tensor in its slow evolution,
we have also presented two other models (II, IV) with non-vanishing
Weyl tensor. Model II, as model I has vanishing radial pressure, whereas
model IV has the same energy density
distribution as model III. This will allow to see the effect of the three
abovementioned factors (local anisotropy, Weyl tensor,
density contrast) on the Tolman-Whittaker mass distribution within the
sphere, and on the velocity profile of different pieces of
matter. With this purpose, all models are considered with the same total
mass $M$ and surface velocity $\omega_\Sigma$.

 Figure 1 exhibits the evolution of $m_{TW}/M$ as function of $x$,
in the process of slow contraction (increasing $n$), for the model I.
For all other models the behaviour is qualitatively the same, i.e. as
the contraction proceeds, the active gravitational mass within
the sphere decreases.
However, the absolute value of $m_{TW}$ is different for
different models (for same $n$ and $x$) as can be seen from figures 2--4,
which display the ratio $m_{TW}/m_{TW}({\rm II})$ for the three models (I,
III, IV).

\vbox{}

As it can be seen, for any $r < r_\Sigma$ (for the same total mass $M$),
we have
$$
m_{TW}({\rm III}) > m_{TW}({\rm I}) > m_{TW}({\rm IV}) > m_{TW}({\rm II})
$$
the differences being larger for more compact (larger $n$) configurations.
Parenthetically, the two conformally flat models present the largest
$TW$ masses.

For models II, III and IV the collapse proceeds in a
quasi-homologous (quasi-linear) regime as indicated in figure 5 for model III
(for models II and IV the figures are similar), deviating from that
regime as $n$ increases.

 However for model I, the contraction is not
homologous even for small $n$ as indicated in figure 6.

Figures 7--9 display the ratio ${\omega}/{\omega({\rm II})}$ for the
three models
(I, III, IV).

 Except for extremely high fields in model I, we see that
$$
\omega({\rm IV}) > \omega({\rm III}) > \omega({\rm II}) > \omega({\rm I})
$$
which indicate that for the same energy density distribution (III,IV) or
radial pressure
distribution (I,II), the slow contraction of interior shells proceeds slower
in the conformally flat case.

\section*{Acknowledgment}
This work was partially supported by the Spanish Ministry of Education
under Grant No. PB96-1306.
We also wish to thank J. L. Hern\'andez-Pastora for his valuable help in
editing this manuscript.

\newpage

\section{Figure captions}

\noindent Figure 1: $m_{TW}/M$ as function of $x$ for the model I, and ten values of
$n$, from .04 to .4.

\vspace{5mm}

\noindent Figure 2: $m_{TW}(I)/m_{TW}(II)$ as function of $x$ ,curves $a-j$ correspond
to $n$=.4; .36; .32; .28;
.24; .20; .16; .12, .08; .04.

\vspace{5mm}

\noindent Figure 3:$m_{TW}(III)/m_{TW}(II)$ as function of $x$ ,for the same values
of $n$ as in figure 2.

\vspace{5mm}

\noindent Figure 4: $m_{TW}(IV)/m_{TW}(II)$ as function of $x$ ,for the same values of
$n$ as in figure 2.

\vspace{5mm}

\noindent Figure 5: ${\omega}/{\omega_\Sigma}$ as function of $x$ for the model III,
and ten values of $n$, from.04 to .4

\vspace{5mm}

\noindent Figure 6: Same as figure 5 for the model I.

\vspace{5mm}

\noindent Figure 7: The ratio $\omega({\rm I})/\omega({\rm II})$ as function of $x$,
for ten values of $n$ from .04 to .4.

\vspace{5mm}

\noindent Figure 8: The ratio $\omega({\rm IV})/\omega({\rm II})$ as function of
$x$,curves $a-j$ correspond to $n$=.4; .36; .32; .28;
.24; .20; .16; .12, .08; .04.

\vspace{5mm}

\noindent Figure 9: The ratio $\omega({\rm III})/\omega({\rm II})$ as function of $x$,
for ten values of $n$ from .04 to .4.

\newpage

\bf{Figure 1}
\begin{figure}[h]
\epsfig{figure=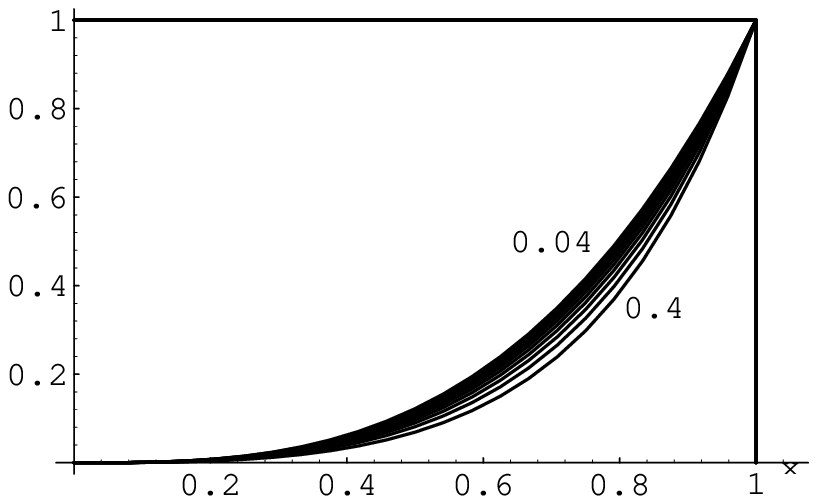,height=3.59in}
\end{figure}

\newpage

\bf{Figure 2}

\begin{figure}[h]
\epsfig{figure=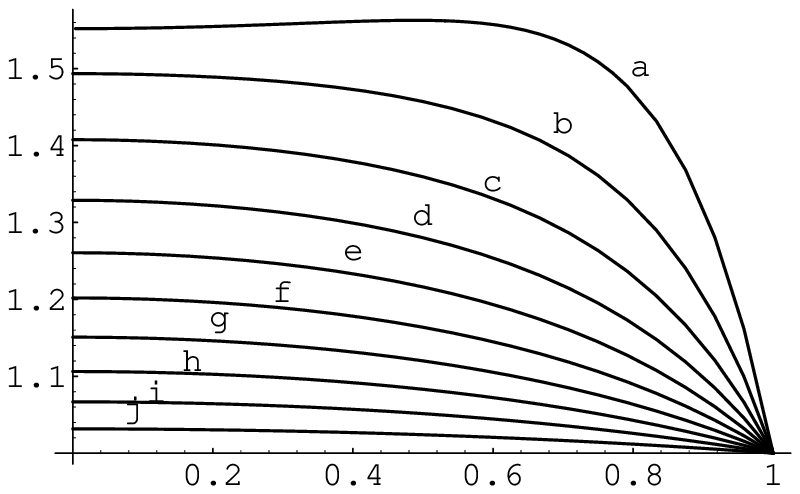,height=3.59in}
\end{figure}

\newpage

\bf{Figure 3}

\begin{figure}[h]
\epsfig{figure=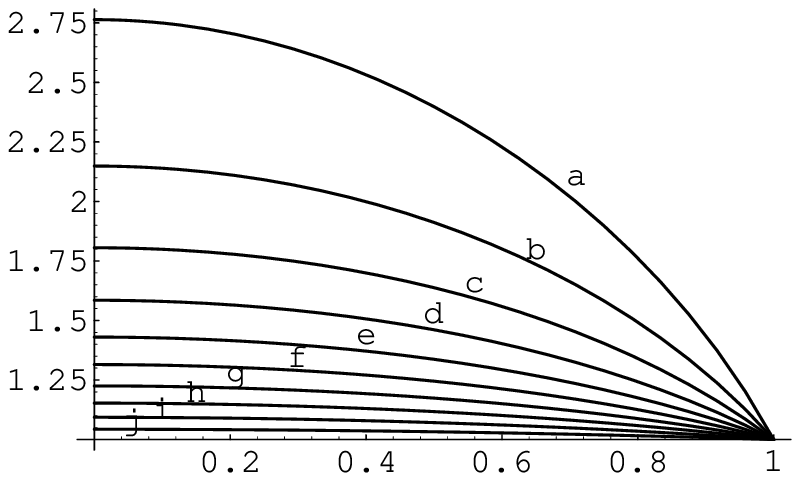,height=3.59in}
\end{figure}

\newpage

\bf{Figure 4}

\begin{figure}[h]
\epsfig{figure=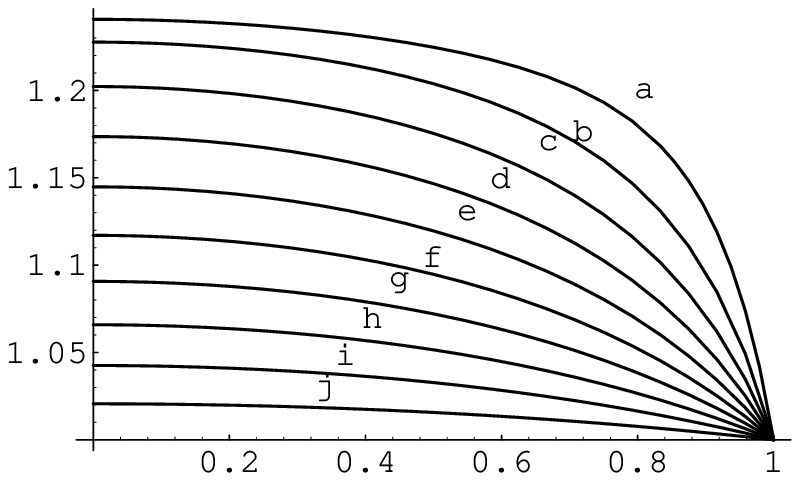,height=3.59in}
\end{figure}

\newpage

\bf{Figure 5}

\begin{figure}[h]
\epsfig{figure=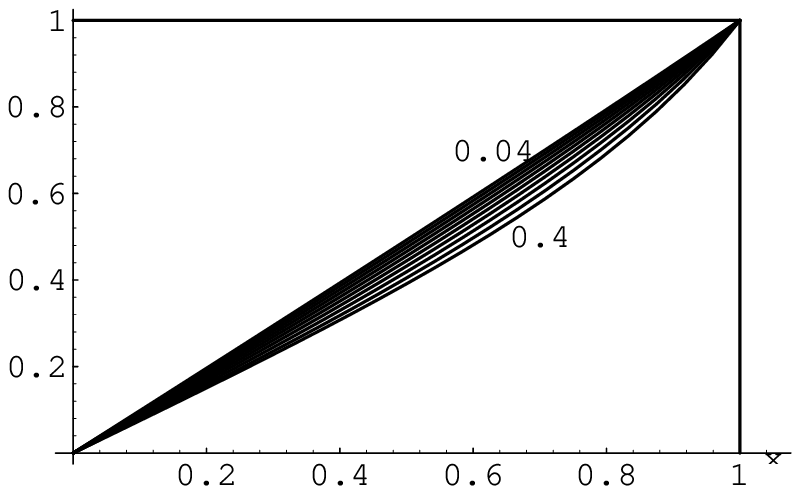,height=3.59in}
\end{figure}

\newpage

\bf{Figure 6}

\begin{figure}[h]
\epsfig{figure=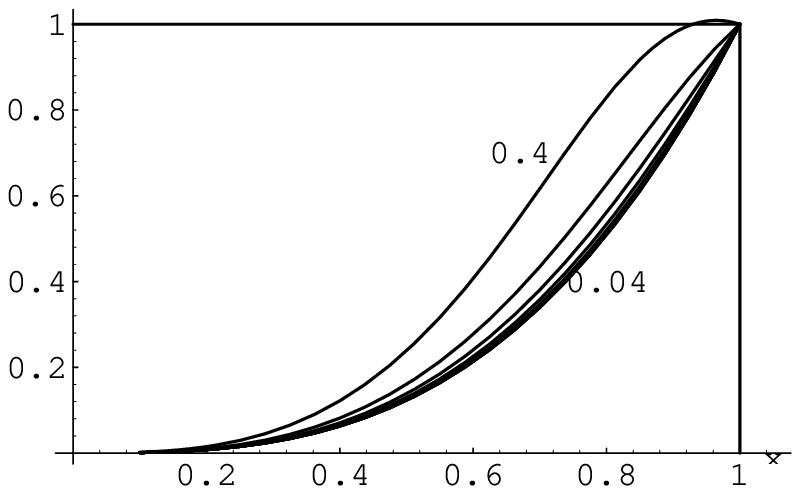,height=3.59in}
\end{figure}

\newpage

\bf{Figure 7}

\begin{figure}[h]
\epsfig{figure=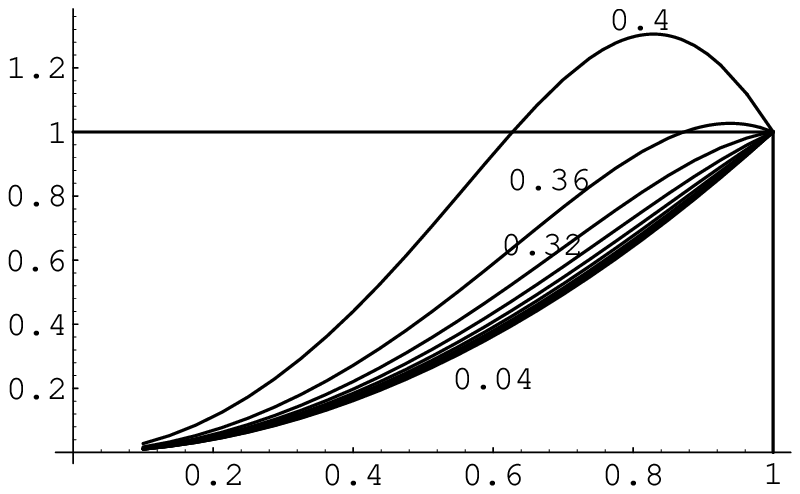,height=3.59in}
\end{figure}

\newpage

\bf{Figure 8}

\begin{figure}[h]
\epsfig{figure=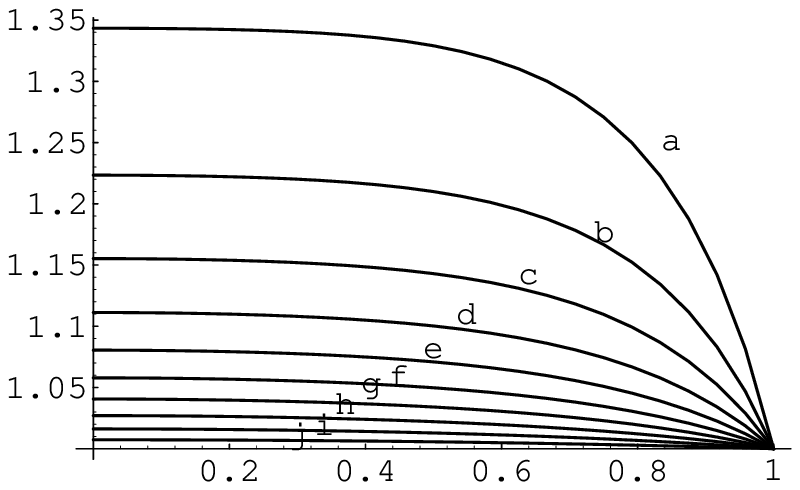,height=3.59in}
\end{figure}

\newpage

\bf{Figure 9}

\begin{figure}[h]
\epsfig{figure=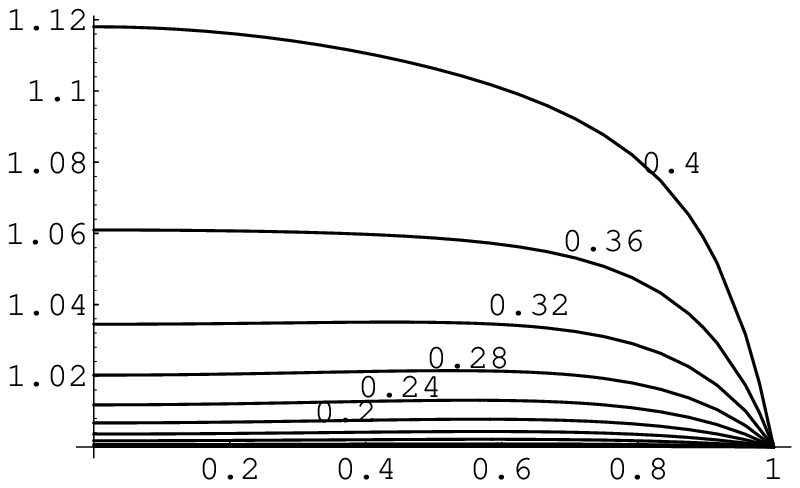,height=3.59in}
\end{figure}

\end{document}